\documentclass[leqno,11pt]{amsart}
\usepackage{amsfonts,latexsym}
\usepackage{amsmath}
\usepackage{amscd}
\usepackage{float,amsmath,amssymb,mathrsfs,bm,multirow,graphics}
\usepackage[dvips]{graphicx}
\usepackage[all]{xy}

\newcommand{\nequation}{\setcounter{equation}{0}}

\newcommand{\R}{{\Bbb R}}

\newcommand{\Z}{{\Bbb Z}}

\newtheorem{theorem}{Theorem}[section]
\newtheorem{proposition}[theorem]{Proposition}

\newtheorem{remark}[theorem]{Remark}

\newtheorem{figuretext}{Figure}

\input epsf
\date{\today}
\title[nonlinear pulse propagation in optical fibers]
{Exactly solvable model for nonlinear pulse propagation in optical fibers}

\author{Jonatan Lenells}
\address{J.L.: Department of Applied Mathematics and Theoretical Physics, 
University of Cambridge, Cambridge CB3 0WA, UK}
\email{j.lenells@damtp.cam.ac.uk} 

\thanks{The author thanks Matz Lenells for valuable discussions and acknowledges support from a Marie Curie Intra-European Fellowship.}
    
\begin{document}
\begin{abstract}
The nonlinear Schr\"odinger (NLS) equation is a fundamental model for the nonlinear propagation of light pulses in optical fibers. We consider an integrable generalization of the NLS equation which was first derived by means of bi-Hamiltonian methods in [A. S. Fokas, {\it Phys. D} {\bf 87} (1995), 145--150]. The purpose of the present paper is threefold: (a) We show how this generalized NLS equation arises as a model for nonlinear pulse propagation in monomode optical fibers when certain higher-order nonlinear effects are taken into account; (b) We show that the equation is equivalent, up to a simple change of variables, to the first negative member of the integrable hierarchy associated with the derivative nonlinear Schr\"odinger equation; (c) We analyze traveling-wave solutions.
\end{abstract}

\maketitle

\noindent
{\small{\sc PACS numbers (2008)}: 42.81.Dp, 02.30.Ik.}


\section{Introduction} \nequation
The nonlinear Schr\"odinger (NLS) equation
\begin{equation}\label{NLS}  
  iu_t + \gamma u_{xx} + \rho |u|^2 u = 0, \qquad x \in \R, \, t > 0,
\end{equation}
where $\gamma$ and $\rho$ are real parameters and $u(x,t)$ is a complex-valued function, is a fundamental model for the nonlinear propagation of light pulses in optical fibers \cite{Agrawal}. Equation (\ref{NLS}) is a completely integrable system and can be linearized by the inverse scattering transform formalism \cite{Z-S}. In the context of fiber optics, the soliton solutions of (\ref{NLS}) are of particular interest: because of their inherent stability, solitons can travel unchanged over long distances, making them ideal carriers of information in optical transmission systems. In this paper we consider a generalization of the NLS equation, which also admits soliton solutions, and which, as we shall argue, models the propagation of nonlinear light pulses in optical fibers when certain higher-order nonlinear effects are taken into account.

The equation
\begin{equation}\label{GNLS}  
  iu_t - \nu u_{tx} + \gamma u_{xx} + \rho |u|^2(u  + i \nu u_x) = 0, \qquad x \in \R, \, t > 0,
\end{equation}
where $\nu, \gamma, \rho$ are real parameters and $u(x,t)$ is a complex-valued function, was derived in \cite{F} as an equation with two distinct, but compatible, Hamiltonian formulations. Just like the bi-Hamiltonian structure of the well-known Korteweg-de Vries equation can be perturbed to yield the integrable so-called Camassa-Holm equation \cite{F-F}, the same mathematical trick applied to the two Hamiltonian operators associated with the NLS equation yields equation (\ref{GNLS})---note that (\ref{GNLS}) reduces to (\ref{NLS}) when $\nu = 0$. Equation (\ref{GNLS}) admits a Lax pair formulation and the initial value problem can be solved by means of the inverse scattering transform \cite{L-F}. 

For large classes of dispersive nonlinear PDE's, the NLS equation arises asymptotically as the equation satisfied at first approximation by the amplitude $A$ of a slowly modulated monochromatic wave \cite{Calogero}. It was demonstrated in \cite{F} that an application of asymptotic techniques to a KdV type equation gives rise to an equation of the form (\ref{GNLS}) provided that one retains terms of the next asymptotic order beyond those included in the case of the NLS equation. In this sense equation (\ref{GNLS}) is a generalization of the NLS equation. 
However, to the best of our knowledge a more direct physical derivation of (\ref{GNLS}) has not yet been presented.

The purpose of the present paper is threefold: (a) We show how equation (\ref{GNLS}) arises as a model for nonlinear wave propagation in monomode optical fibers; (b) We show that (\ref{GNLS}) is equivalent, up to a simple change of variables, to the first negative member of the integrable hierarchy associated with the derivative nonlinear Schr\"odinger (DNLS) equation; (c) We analyze traveling-wave solutions.

Let us comment on (a), (b), and (c).
\bigskip

\noindent
{\bf (a) Physical derivation.}
The propagation of nonlinear pulses in optical fibers is described to first order by the NLS equation (\ref{NLS}). However, when considering very short input pulses it is necessary to include additional terms accounting for higher-order effects such as self-steepening, Raman scattering, and third-order dispersion \cite{Kodama1985, K-H}. While the rmore general esulting equation is not integrable, there exist exactly solvable approximations which capture various aspects of the higher-order phenomena. Most notably, the DNLS, Hirota, and Sasa-Satsuma equations are all integrable reductions of the more general equation and are used as models when studying wave propagation for ultrashort input pulses \cite{A-L, Kodama1985, T-J}. 
Although each of these reductions ignores some of the important nonlinear effects, their solutions can be used as a first approximation when performing a more complete analysis involving perturbation methods. Moreover, the possibility of presenting exact solutions to these models make them valuable tools for understanding the physical influence of the various terms. 

After recalling certain aspects of the standard derivation of the NLS equation in nonlinear fiber optics, we indicate in Section \ref{physicalsec} how equation (\ref{GNLS}) appears when taking into account terms that are normally ignored. 
\bigskip

\noindent
{\bf (b) Relation to DNLS.}
The derivative nonlinear Schr\"odinger equation 
\begin{equation}\label{DNLS}
  iq_t + q_{xx} + \sigma i(|q|^2q)_x = 0, \qquad \sigma = \pm 1,
\end{equation}
and gauge transformed versions thereof, have several applications in plasma physics and nonlinear fiber optics \cite{Mjolhus76, T-J}. It was shown in \cite{K-N} that (\ref{DNLS}) admits a Lax pair formulation and can be solved by means of inverse scattering techniques. Being integrable, equation (\ref{DNLS}) admits an infinite number of conservation laws and a bi-Hamiltonian formulation.
Moreover, associated with equation (\ref{DNLS}) is an infinite hierarchy of equations generated by the bi-Hamiltonian structure, equation (\ref{DNLS}) being the second positive member. In Subsection \ref{bihamsec} we demonstrate that, up to a simple change of variables, equation (\ref{GNLS}) is nothing but the first negative member of this hierarchy given by
\begin{equation}\label{GNLSgaugeintro}
  u_{tx} =  u - i\sigma |u|^2 u_{x}, \qquad \sigma = \pm 1.
\end{equation}  
Since the parameter $\sigma$ in (\ref{GNLSgaugeintro}) can be removed by the replacement $(t,x) \to (\sigma t,  \sigma x)$, this shows that the parameters $\nu, \gamma$, $\rho$ in equation (\ref{GNLS}) can all be set to $1$ by a change of variables. In particular, in contrast to the case of the NLS equation which comes in a focusing as well as in a defocusing version depending on the values of the parameters, and solitons only exist in the the focusing regime, all versions of equation (\ref{GNLS}) are mathematically equivalent up to a change of variables. This observation is consistent with the fact that solitons for (\ref{GNLS}) exist for all values of $\nu, \gamma$, $\rho$ \cite{L-F}.

Our investigation of the relationship between (\ref{GNLS}) and (\ref{DNLS}) was motivated by the observation that if $u_x$ is identified with $q$, then the Lax pairs associated with equations (\ref{GNLS}) and (\ref{DNLS}) have identical $x$-parts \cite{L-F}. Let us point out that a similar link exists between the KdV and Camassa-Holm equations: The Camassa-Holm equation is derived mathematically from the KdV equation in an analogous manner that (\ref{GNLS}) is derived from the NLS equation. Moreover, it is known that the Camassa-Holm equation is related by a (nonlocal) change of variables to the first negative member of the KdV hierarchy \cite{McK}. Since the gauge transformation relating (\ref{GNLS}) to (\ref{DNLS}) is a local change of variables, the link in the present case is more direct.
\bigskip

\noindent
{\bf (c) Traveling waves.}
In Section \ref{travsec} we consider traveling-wave solutions of (\ref{GNLS}). For simplicity, we consider the simpler, but equivalent, equation (\ref{GNLSgaugeintro}). If we assume that the solution $u(x,t)$ of (\ref{GNLSgaugeintro}) has the special form
\begin{equation}\label{travform}  
  u(x,t) = \varphi(x-ct)e^{i\left(kx - \Omega t + \theta(x - ct)\right)},
\end{equation}
where $k$, $\Omega$, $c$ are real parameters and $\varphi$ and $\theta$ are real-valued functions, the partial differential equation for $u$ reduces to a first-order ODE for $\varphi$. A straightforward analysis of this equation yields the qualitative structure of large classes of traveling-wave solutions. 
For certain values of the parameters the equation for $\varphi$ can be integrated explicitly. In particular, we recover the one-soliton solutions earlier found by means of the inverse scattering approach \cite{L-F}. 
We also comment on the occurence of more exotic traveling waves such as peakons. A peakon is a solution whose profile has a peak at its crest, see Figure \ref{peakon.pdf}. The existence of peaked traveling-wave solutions is a well-known property of the Camassa-Holm equation \cite{C-H}. Since (\ref{GNLS}) is related to the NLS equation by a procedure analogous to that which gives the Camassa-Holm equation from KdV, it is natural to investigate whether (\ref{GNLS}) also exhibits weak solutions of this kind. Our discussion in Section \ref{travsec} suggests that (\ref{GNLS}) admits no peaked traveling waves which are weak solutions in any reasonable sense.

\section{Physical derivation}\label{physicalsec}\nequation
Consider the propagation of an optical pulse in a monomode fiber aligned in the $z$-direction with a frequency-dependent dielectric constant. We first follow the standard derivation of the NLS equation (see e.g. \cite{Agrawal}), before we indicate how the additional higher-order effects particular to equation (\ref{GNLS}) arise. Maxwell's equations lead to the basic equation 
 \begin{equation}\label{nablanablaE} 
 \nabla \times \nabla \times \mathbf{E} = -\frac{1}{c^2} \mathbf{E}_{tt} - \mu_0 \mathbf{P}_{tt},
\end{equation}
where $\mathbf{E}(\mathbf{r}, t)$ and $\mathbf{P}(\mathbf{r}, t)$ are the electric field and the induced polarization evaluated at time $t$ at the spatial point $\mathbf{r}$, respectively, $\mu_0$ is the vacuum permeability, $c$ is the speed of light in vacuum, and subscripts denote partial derivatives.
Assuming a local medium response and including only third-order nonlinear effects, the total polarization $\mathbf{P}(\mathbf{r}, t) = \mathbf{P}_L(\mathbf{r}, t) + \mathbf{P}_{NL}(\mathbf{r}, t)$ can be split into a linear part
$$\mathbf{P}_L = \epsilon_0 \int_{-\infty}^t \chi^{(1)}(t - t') \mathbf{E}(\mathbf{r}, t')dt',$$
and a nonlinear part 
$$\mathbf{P}_{NL} = \epsilon_0 \int_{-\infty}^t dt_1\int_{-\infty}^t dt_2\int_{-\infty}^t dt_3 \chi^{(3)}(t - t_1, t - t_2, t-t_3)[\mathbf{E}(\mathbf{r}, t_1), \mathbf{E}(\mathbf{r}, t_2), \mathbf{E}(\mathbf{r}, t_3)]dt',$$ 
where $\epsilon_0$ is the vacuum permittivity and $\chi^{(j)}$ is the $j$th-order susceptibility. 
If we moreover assume that the nonlinear response of the fiber is instantaneous, we find
\begin{equation}\label{PNLinstant} 
 \mathbf{P}_{NL}(\mathbf{r}, t) = \epsilon_0 \chi^{(3)}[\mathbf{E}(\mathbf{r}, t), \mathbf{E}(\mathbf{r}, t), \mathbf{E}(\mathbf{r}, t)],
\end{equation}
where $\chi^{(3)}$ is a time-independent map linear in its three arguments. Since we consider a quasi-monochromatic pulse which maintains its polarization along the fiber length, a scalar approach is appropriate. If the pulse spectrum is centered at $\omega_0$, we write
\begin{align}\label{EPLPNL} \nonumber
& \mathbf{E}(\mathbf{r},t) = \frac{\mathbf{\hat{x}}}{2}\left(E(\mathbf{r},t) e^{-i\omega_0 t} +c.c.\right),
	\\
& \mathbf{P}_L(\mathbf{r},t) = \frac{\mathbf{\hat{x}}}{2}\left(P_L(\mathbf{r},t) e^{-i\omega_0 t} +c.c.\right),
	\\ \nonumber
& \mathbf{P}_{NL}(\mathbf{r},t) = \frac{\mathbf{\hat{x}}}{2}\left(P_{NL}(\mathbf{r},t) e^{-i\omega_0 t} +c.c.\right),
\end{align}
where $\mathbf{\hat{x}}$ is a unit vector pointing in the direction of the polarization, c.c. stands for the complex conjugate, and $E$, $P_L$, $P_{NL}$ are complex-valued slowly-varying functions. Substituting (\ref{EPLPNL}) into (\ref{PNLinstant}) and ignoring terms of frequency $3 \omega_0$, we deduce that $P_{NL} =  \epsilon_0 \epsilon_{NL} E$, where $\epsilon_{NL} = \frac{3}{4} \chi_{xxxx}^{(3)}|E(\mathbf{r}, t)|^2$ is the nonlinear contribution to the dielectric constant responsible for the Kerr effect ($\chi_{xxxx}^{(3)}$ denotes the $xxxx$-component of $\chi^{(3)}$).
Taking the Fourier transform of equation (\ref{nablanablaE}) while making the approximate assumption that $\epsilon_{NL}$ is a constant, we obtain
\begin{equation}\label{Helmholtz}  
  \nabla^2 \tilde{E}(\mathbf{r}, \omega - \omega_0) + \epsilon(\omega) k_0^2 \tilde{E}(\mathbf{r}, \omega - \omega_0) = 0,
\end{equation}
where $k_0 = \omega/c$, the Fourier transform $\tilde{E}$ is defined by
$$\tilde{E}(\mathbf{r}, \omega - \omega_0) = \int_\R E(\mathbf{r},t) e^{i(\omega - \omega_0) t} dt,$$
and the frequency-dependent dielectric constant
$$\epsilon(\omega) = 1 + \tilde{\chi}_{xx}^{(1)}(\omega) + \epsilon_{NL},$$
contains the Fourier transform $\tilde{\chi}_{xx}^{(1)}(\omega)$ of the $xx$-component of $\chi^{(1)}$.

We seek a solution to (\ref{Helmholtz}) of the form
$$\tilde{E}(\mathbf{r},\omega - \omega_0) = F(x,y) \tilde{A}(z, \omega - \omega_0) e^{i \beta_0 z},$$
where $\tilde{A}(z, \omega)$ is a slowly varying function of $z$ and $\beta_0$ is the wave number accounting for the fast oscillations. Substituting this into (\ref{Helmholtz}), we find
\begin{align}\label{Feq}
 F_{xx} + F_{yy} + \left(\epsilon(\omega) k_0^2 - \tilde{\beta}(\omega)^2\right) F = 0,
 	\\ \label{tildeAeq}
\tilde{A}_{zz} + 2 i \beta_0 \tilde{A}_z + \left(\tilde{\beta}(\omega)^2 - \beta_0^2\right) \tilde{A} = 0,
\end{align}
where $\tilde{\beta}(\omega)$ is a function independent of $\mathbf{r}$.
At this point we have deviated from the standard derivation in which the $\tilde{A}_{zz}$ term in equation (\ref{tildeAeq}) is ignored since $A(z, t)$ is assumed to describe the slowly varying envelope of the pulse. 

The derivation proceeds by solving (\ref{tildeAeq}) using perturbation theory and 
then applying the inverse Fourier transform in order to return to the time domain. The result is (see \cite{Agrawal} for details in the case when $\tilde{A}_{zz}$ is ignored)
\begin{equation}\label{Azzeq}
  - \frac{i}{2 \beta_0}A_{zz} + A_z + \beta_1 A_t + \frac{i \beta_2}{2}A_{tt} + \frac{\alpha}{2} A = i  \rho |A|^2 A,
\end{equation}
where the terms proportional to $\alpha$ and $\rho$ include the effects of fiber loss and nonlinearity, respectively. The coefficients $\beta_0, \beta_1, \beta_2$ are
determined by the expansion of $\beta(\omega) = n(\omega)\omega/c$, where $n(\omega)$ is the refractive index, around the carrier frequency $\omega_0$ according to
$$\beta(\omega) = \beta_0 + \beta_1(\omega - \omega_0) + \frac{\beta_2}{2}(\omega - \omega_0)^2 + \frac{\beta_3}{3!}(\omega - \omega_0)^3 + \cdots.$$
In the frame of reference traveling with the pulse at the group velocity $v_g = 1/\beta_1$, equation (\ref{Azzeq}) becomes
\begin{equation}\label{iAzAzz}
i A_z + \frac{1}{2\beta_0} A_{zz} - \frac{1}{\beta_0 v_g} A_{zT} + \gamma A_{TT} 
+ \frac{i \alpha}{2} A = -  \rho |A|^2A,
\end{equation}
where $\gamma = \frac{1}{2\beta_0 v_g^2} - \frac{\beta_2}{2}$ and we have introduced the new variable $T$ by $T = t - \beta_1 z$.

As compared with the standard derivation, we see that our retaining of $\tilde{A}_{zz}$ in equation (\ref{tildeAeq}) has given rise to two additional terms in (\ref{iAzAzz}) involving $A_{zz}$ and $A_{zT}$. If these are ignored and the dissipation is set to zero (i.e. $\alpha = 0$), equation (\ref{iAzAzz}) reduces to the NLS equation. 

The terms proportional to $A_{zz}$ and $A_{zT}$ are usually ignored due to the inequalities (see \cite{H-W})
$$|A_{zz}| \ll \beta_0 |A_z|, \qquad |A_{zT}| \ll \beta_0 v_g |A_z|,$$
assumed to be valid if $A$ is a slowly varying envelope. However, in the (femtosecond) regime of very short pulses, the pulse envelope $A$ may contain only a few optical cycles. Hence these inequalities are expected to be less strong in this range and it motivates us to not throw away the terms involving $A_{zz}$ and $A_{zT}$. On the other hand, for very short pulses the approximation (\ref{PNLinstant}) also has to be corrected in order to incorporate higher-order nonlinear effects. A more careful analysis \cite{K-H} taking additional effects into account shows that the right-hand side of equation (\ref{iAzAzz}) should be replaced by 
$$\frac{i \beta_3}{6}  A_{TTT} -  \rho A |A|^2  - i s \left(A|A|^2\right)_T - i \tau A \left(|A|^2\right)_T,$$
where $\beta_3$ governs the effect of third-order dispersion, $s$ accounts for the so-called self-steepening effect, and $\tau$ is in general a complex parameter arising from the retarded nonlinear reponse of the medium; in particular, the imaginary part of $\tau$ describes the retarded Raman effect in which a photon through scattering pass on part of its energy to a vibrational mode of a molecule in the medium. The Raman effect is a nonlinear dissipative effect which causes a frequency downshift of the pulse \cite{Agrawal, K-H}.

While our final equation
\begin{align}\label{finaleq}
  i A_z + \frac{1}{2\beta_0} A_{zz} - \frac{1}{\beta_0 v_g} A_{zT} &+  \gamma A_{TT} + \frac{i \alpha}{2} A
- \frac{i \beta_3}{6} A_{TTT} 
	\\ \nonumber
& = -  \rho A |A|^2  - i s \left(A|A|^2\right)_T - i \tau A \left(|A|^2\right)_T
\end{align}
is rather complicated, there exist some special choices of the parameters for which the equation is integrable and hence exactly solvable by means of the inverse scattering formalism. First of all we need to set the linear and nonlinear dissipative terms to zero (i.e. $\alpha = 0$ and $\text{Im}\, \tau = 0$) in order to get an energy-conserving system. Then several integrable reductions exist.

\begin{itemize}
\item When the terms proportional to $A_{zz}$, $A_{zT}$, $A_{TTT}$, and $A \left(|A|^2\right)_T$ are neglected, equation (\ref{finaleq}) reduces to
\begin{equation}\label{CLLeq} 
  i A_z +   \gamma A_{TT}   = -  \rho A |A|^2  - i s \left(A|A|^2\right)_T.
\end{equation}
This equation is integrable and admits soliton solutions \cite{C-L-L}. In fact, a gauge transformation converts it into the DNLS equation (\ref{DNLS}).

\item More generally, when the terms proportional to $A_{zz}$ and $A_{zT}$ are ignored, we obtain 
\begin{equation}\label{reduced}  
  i A_z +   \gamma A_{TT}  - \frac{i \beta_3}{6} A_{TTT} = -  \rho A |A|^2  - i s \left(A|A|^2\right)_T - i \tau A \left(|A|^2\right)_T.
\end{equation}
There are two known \cite{Sak} integrable subcases of this equation apart from (\ref{CLLeq}); both occur when the parameters satisfy $\beta_3 \rho = -2 s \gamma$. When this condition is fulfilled, a gauge transformation brings (\ref{reduced}) to the form \cite{GHNO}
\begin{equation}\label{reduced2}
  i A_z  - \frac{i \beta_3}{6} A_{TTT} = - i s |A|^2A_T - i(s + \tau)A \left(|A|^2\right)_T.
\end{equation}
If the ratios of the coefficients $ - \frac{\beta_3}{6} $, $s$, and $s + \tau$ in (\ref{reduced2}) are related as $1:6:0$ we obtain Hirota's equation \cite{Hirota}
$$A_z + A_{TTT} = - 6|A|^2A_T,$$
while if they are related as $1:6:3$ we find the Sasa-Satsuma equation \cite{S-S}
$$A_z + A_{TTT} = - 6|A|^2A_T - 3A \left(|A|^2\right)_T.$$

\item When the terms proportional to $A_{zz}$ and $A_{TTT}$ are ignored and $s + \tau = 0$, equation (\ref{finaleq}) becomes
\begin{equation}\label{GNLSphysical}
 i A_z - \frac{1}{\beta_0 v_g} A_{zT} +  \gamma A_{TT} = -  \rho |A|^2\left(A  + i \frac{s}{ \rho} A_T\right),
 \end{equation}
which is equation (\ref{GNLS}) if $\frac{1}{\beta_0 v_g}  = \frac{s}{ \rho}$.\footnote{The variables $A$, $z$, and $T$ in (\ref{GNLSphysical}) are identified with $u$, $t$, and $x$ in (\ref{GNLS}), respectively. The parameter $\nu$ is identified with $\frac{1}{\beta_0 v_g} = \frac{s}{ \rho}$.}

\end{itemize}

We conclude that these four exactly solvable models arise as reductions of the full nonlinear equation (\ref{finaleq}) when different choices are made as to which higher-order effects to include. Although none of them can be expected to exhibit all the structure tied into (\ref{finaleq}) (e.g. they all ignore the effects of fiber loss), they do provide important first approximations to the full theory. 
 In this sense equation (\ref{GNLS}) can be viewed as a model for light pulses in single-mode optical fibers. 

\section{Relation to DNLS} \nequation
It was found in \cite{L-F} that the $x$-part of the Lax pair of (\ref{GNLS}) is simply related to the $x$-part of the Lax pair of the derivative nonlinear Schr\"odinger equation (\ref{DNLS}). More precisely, the two $x$-parts are identical if $u_x$ is identified with $q$, where $u$ is the solution of (\ref{GNLS}) and $q$ is the solution of (\ref{DNLS}). Although the corresponding $t$-parts are very different, this suggests a deep relationship between (\ref{GNLS}) and (\ref{DNLS}). Since all members of the hierarchy associated with an integrable equation admit the same $x$-part, it is reasonable to expect the equation describing the evolution of $u_x$ according to (\ref{GNLS}) to be related to a member of the DNLS hierarchy. We will show that a simple change of variables indeed transforms (\ref{GNLS}) into the first negative member of this hierarchy.

\subsection{Gauge transformation}\label{gaugesec}
Our first step is to transform equation (\ref{GNLS}) by means of a gauge transformation.
Replacing $x$ by $-x$ if necessary, we can assume that the parameters $\gamma$ and $\nu$ in (\ref{GNLS}) have the same sign. Then, letting $a = \gamma/\nu >0$ and $b = 1/\nu$, the gauge tranformation
$$u \to \sqrt{\frac{a}{|\rho|}}be^{i(bx + 2abt)}u,$$
transforms (\ref{GNLS}) into
$$u_{tx} - a u_{xx} = ab^2(- u + i\sigma |u|^2 u_x), \qquad \sigma = \hbox{sgn}\, \rho = \pm 1.$$
Introducing new variables by 
$$\xi = x + a t, \qquad \tau = -ab^2 t,$$
we arrive at the equation
\begin{equation}\label{GNLSgauge}
  u_{\tau \xi} =  u - i\sigma |u|^2 u_{\xi}, \qquad \sigma = \pm 1,
\end{equation}  
which is exactly (\ref{GNLSgaugeintro}) up to a relabeling of $\tau$ and $\xi$. Putting the various transformations together we arrive at the following result.

\begin{proposition} Fix any nonzero values of the parameters $\nu, \gamma, \rho \in  \R$. Then $u(x,t)$ satisfies equation (\ref{GNLS}) if and only if $u_g(\xi, \tau)$ satisfies equation (\ref{GNLSgauge}) with $\sigma = \text{\upshape sgn}(\nu \gamma \rho)$ where $u$ and $u_g$ are related as
$$u(x,t) = \sqrt{\text{\upshape sgn}(\nu \gamma \rho) \frac{\gamma}{\nu^3 \rho}} e^{i\left(\frac{x}{\nu} + \frac{2\gamma t}{\nu^2}\right)} u_g\left(x + \frac{\gamma}{\nu}t, -\frac{\gamma}{\nu^3} t\right).$$
\end{proposition}

\subsection{Bi-Hamiltonian structure}\label{bihamsec}
Equations (\ref{DNLS}) and (\ref{GNLSgaugeintro}) are equivalent to the systems
\begin{equation}\label{DNLSsystem}
\begin{pmatrix} q \\ r \end{pmatrix}_t
=  \begin{pmatrix}  iq_{xx} - (q^2r)_x \\
 -ir_{xx} - (qr^2)_x \end{pmatrix}
 \end{equation}
and
\begin{equation}\label{GNLSgaugesystem}
 \begin{pmatrix} u_x \\ v_x \end{pmatrix}_t
=  \begin{pmatrix}  u - iuvu_x \\
 v + iuvv_x \end{pmatrix},
\end{equation}
respectively, when $r = \sigma \bar{q}$ and $v = \sigma \bar{u}$.
The operators 
$$J_1 =  \begin{pmatrix} 
 0	&	\partial_x	\\
 \partial_x		&	0 \end{pmatrix}, \qquad J_2 = \begin{pmatrix} 
 - q \partial_x^{-1}q &  i +  q \partial_x^{-1} r 
 \\  - i + r \partial_x^{-1} q   & - r \partial_x^{-1} r \end{pmatrix},$$
form a compatible pair of Hamiltonian operators cf. \cite{M-Z}. 
Starting with the Hamiltonian
$$H_0 = -\int qr dx,$$
the operators $J_1$ and $J_2$ generate a hierarchy of bi-Hamiltonian equations according to\footnote{The gradient of a functional $F[q,r]$ is defined by
$$\text{grad}\, F = \begin{pmatrix} \frac{\delta F}{\delta q} \\ \frac{\delta F}{\delta r} \end{pmatrix},$$
whenever there exist functions $\frac{\delta F}{\delta q}$ and $\frac{\delta F}{\delta r}$ such that, for any smooth functions $\varphi_1$ and $\varphi_2$,
$$\frac{d}{d\epsilon} F[q + \epsilon \varphi_1, r + \epsilon \varphi_2]\biggr|_{\epsilon =0} = \int \biggl(\frac{\delta F}{\delta q} \varphi_1 + \frac{\delta F}{\delta r} \varphi_2 \biggr)dx.$$}
$$\begin{pmatrix} q \\ r \end{pmatrix}_t = K_n = J_1 \text{grad}\, H_{n-1} = J_2 \text{grad}\, H_n, \qquad n \in \Z,$$
where the infinite sequence of conservation laws $\{H_n\}_{n \in \Z}$ are constructed recursively from the relation
$$J_1 \text{grad}\, H_n = J_2 \text{grad}\, H_{n+1}, \qquad n \in \Z.$$
The first few members of this hierarchy are presented in Figure \ref{hierarchyfig}, where we have written $u$ and $v$ for $\partial_x^{-1}q$ and $\partial_x^{-1}r$, respectively.
We immediately recognize equation (\ref{DNLSsystem}) as the second member and equation (\ref{GNLSgaugesystem}) as the first negative member of this hierarchy. 


 \begin{figure}
\begin{center}
$$ \xymatrix{
  &  \ar[dl]_{J_2}   \text{grad}\, \int \left(q r_{xx}-\frac{1}{2} q^3 r^3-\frac{3i}{2} q^2 rr_{x}\right) dx = \text{grad}\, H_2	
  		\\
{\begin{pmatrix} q \\ r \end{pmatrix}_t
=  \begin{pmatrix}  iq_{xx} - (q^2r)_x \\
 -ir_{xx} - (qr^2)_x \end{pmatrix}} & & 
 		\\  
    &   \ar[ul]_{J_1} \ar[dl]_{J_2}   \text{grad}\, \int \left(-\frac{1}{2} q^2 r^2-i q r_x\right) dx = \text{grad}\, H_1\\
{\begin{pmatrix} q \\ r \end{pmatrix}_t
= - \begin{pmatrix} q_x  \\
 r_x \end{pmatrix}}  &  
 		\\
  &   \ar[ul]_{J_1} \ar[dl]_{J_2}   \text{grad}\, \int \left(-qr \right) dx = \text{grad}\, H_0
  		\\
{\begin{pmatrix} q \\ r \end{pmatrix}_t
= i \begin{pmatrix} -q \\
 r \end{pmatrix}}   & 
		 \\
 &   \ar[ul]_{J_1}  \ar[dl]_{J_2}    \text{grad}\, \int ivu_x dx = \text{grad}\, H_{-1}	
		 \\
{\begin{pmatrix} u_x \\ v_x \end{pmatrix}_t
= \begin{pmatrix} u - i u vu_x \\
 v + i u vv_x \end{pmatrix}}   &   
 		 \\
 &   \ar[ul]_{J_1}     \text{grad}\, \int \left(-uv + \frac{i}{2}uv^2u_x\right) dx = \text{grad}\, H_{-2}
		 }$$
     \begin{figuretext}\label{hierarchyfig}
       Recursion scheme for the operators $J_1$ and $J_2$. The notation $u = \partial_x^{-1}q$ and $v = \partial_x^{-1}r$ has been used when convenient.
     \end{figuretext}
     \end{center}
\end{figure}

\begin{remark}
\upshape Since the inverse of $J_2$ is given explicitly by
$$J_2^{-1} = \begin{pmatrix} 
 r \partial_x^{-1}r &  i +  r \partial_x^{-1} q 
 \\  - i + q \partial_x^{-1} r   & q \partial_x^{-1} q \end{pmatrix},$$
it is straightforward to implement the recursive scheme for construction of the $H_n$'s on a computer and thus obtain explicit formulas for as many conserved quantities as one wishes. The $H_n$'s are local in $q$ and $r$ for $n \geq 0$, while they involve the nonlocal operator $\partial_x^{-1}$ for $n < 0$. In both cases, the expressions grow rapidly in size, e.g. $H_3$ and $H_4$ are given by
\begin{align*}
H_3 = \int \biggl(&-\frac{5}{8} q^4 r^4-\frac{5}{2} i q^3 r_xr^2+\frac{1}{2}
   q q_{xx} r^2+q q_x r_x r+2 q^2 r_{xx}r+\frac{3}{2} q^2 r_x^2+i q r_{xxx}\biggr)dx,
   	\\
H_4 = \int \biggl(&-\frac{7}{8} q^5 r^5+\frac{5}{6} q q_x^2 r^3-\frac{35}{8} i
   q^4 r_x r^3+\frac{5}{3} q^2 q_{xx} r^3+5 q^2 q_x
   r_x r^2+\frac{25}{6} q^3 r_{xx} r^2
   	\\
   &+\frac{35}{6} q^3
   r_x^2 r+\frac{5}{2} i q r_x q_{xx} r+\frac{5}{2} i q
   q_x r_{xx} r+\frac{5}{2} i q^2 r_{xxx} r+\frac{5}{2} i
   q q_x r_x^2
   	\\
 &  +5 i q^2 r_x r_{xx}-q r_{xxxx}\biggr)dx.
\end{align*}
\end{remark}

\section{Traveling waves}\nequation\label{travsec}
In this section we analyze the traveling-wave solutions of equation (\ref{GNLS}) taken in the simpler, but equivalent, form (\ref{GNLSgaugeintro}). Letting $(t,x) \to (\sigma t, \sigma x)$, we may assume that $\sigma = 1$ in (\ref{GNLSgaugeintro}). Thus consider the equation
\begin{equation}\label{GNLSgaugesigma1}
  u_{tx} =  u - i |u|^2 u_{x}.
\end{equation}  
Substituting
$$u(x,t) = W(x,t)e^{i\chi(x,t)},$$
with real-valued functions $W$ and $\chi$ into (\ref{GNLSgaugesigma1}), we obtain the coupled equations
\begin{align}\label{Wchicoupled}
\begin{cases}
-\chi_x W^3-\left(\chi_t \chi_x+1\right) W + W_{tx} = 0,
	\\
W_x W^2+\chi_{tx} W + W_t \chi_x + \chi_t W_x = 0.
 \end{cases}
\end{align}
We seek a solution of the form
\begin{equation}\label{chiW}
  \chi(x,t) = kx - \Omega t + \theta(y),  \qquad W(x,t) = \varphi(y),
\end{equation}
where $k$, $\Omega$, $c$ are real parameters and $y = x -c t$. Equations (\ref{Wchicoupled}) and (\ref{chiW}) yield
\begin{align}\label{travcoupled}
\begin{cases}
-\left(k+\theta_y\right) \varphi^3-\left(1-\left(k+\theta_y \right)\left(\Omega+c \theta_y \right)\right) \varphi - c\varphi_{yy} = 0
      	\\ 
 -\varphi_y \left(-\varphi^2+c k+\Omega+2 c \theta_y \right) - c \varphi \theta_{yy} = 0 
  \end{cases}
\end{align}
Multiply the second of these equations by $\varphi$ and integrate the resulting equation to find
$$\frac{\varphi^4}{4} - \frac{1}{2} \left(c k+\Omega + 2 c \theta_y \right) \varphi^2+A = 0,$$
for some constant of integration $A$. Solving for $\theta_y$, we obtain
\begin{equation}\label{thetay}  
  \theta_y = \frac{\varphi^4-2 c k \varphi^2-2 \Omega \varphi^2+4 A}{4 c \varphi^2}.
\end{equation}  
We use this equation to eliminate $\theta_y$ from the first equation in (\ref{travcoupled}). Integration of the resulting equation multiplied by $\varphi_y$ yields
\begin{equation}\label{varphiy2V}  
  \varphi_y^2 = -\frac{1}{16 c^2} V(\varphi),
  \end{equation}
where
\begin{align*}
&V(\varphi) = \frac{\varphi^8+ c_3
   \varphi^6+c_2 \varphi^4+ c_1
   \varphi^2+c_0}{\varphi^2},
   	\\
& c_3 =4 c k-4 \Omega,
	\\
& c_2 =4 \left(c^2 k^2+\Omega^2+2 A+c (4-2 k \Omega)\right),
	\\
& c_1 = -32 B c,
	\\
& c_0 = 16 A^2,
\end{align*}
and $B$ is another integration constant. Solutions of (\ref{varphiy2V}) can only exist in regions where $V(\varphi) \leq 0$. Since $c_0 > 0$ unless $A = 0$, we infer that solitary waves (i.e. waves such that $\varphi$ decays to zero as $y \to \pm \infty$) can exist only when $A = 0$. A classification of the traveling waves can be obtained by analyzing the distribution of the zeros of $V$ as the parameters $\Omega, c, k, A, B$ vary. Rather than completing this program in detail, we simply indicate how the one-soliton solutions arise and discuss the (non-)existence of peaked traveling waves.

\begin{figure}
\begin{center}
    \includegraphics[width=.45\textwidth]{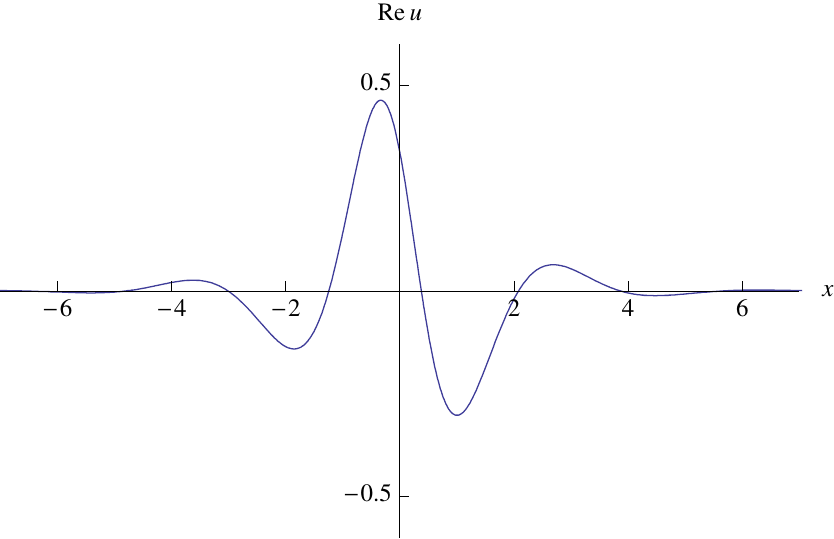} \quad
    \includegraphics[width=.45\textwidth]{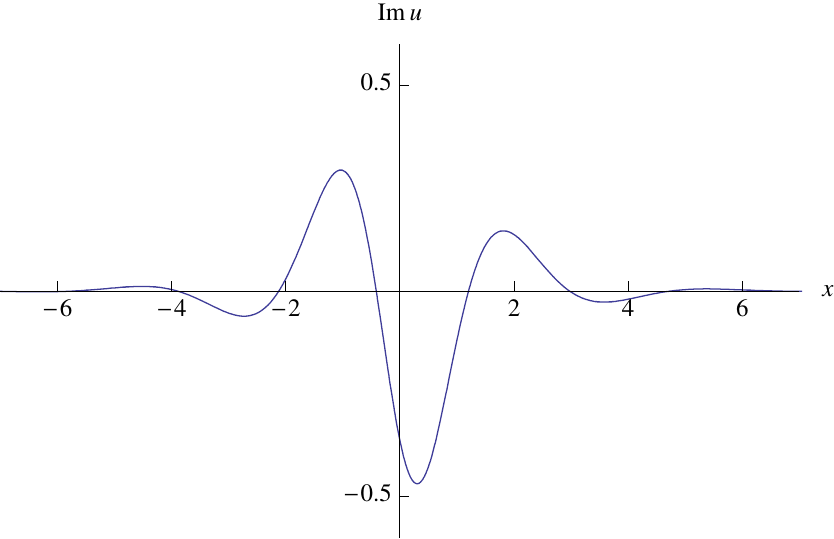} \\
    \includegraphics[width=.5\textwidth]{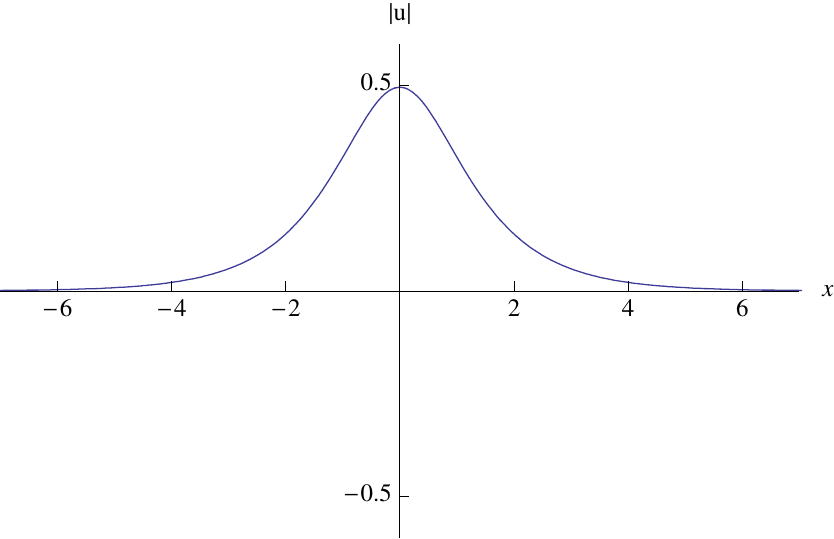} \quad
     \begin{figuretext}\label{solitongraphs}
       The real part, imaginary part, and the absolute value at $t = 0$ of the one-soliton solution $u(x,t)$ given by equation (\ref{onesolitongauge}). The graphs correspond to the parameter-values $\Delta = 1$, $\gamma = 1/2$, and $\Sigma_0 = \Theta_0 = 0$.
     \end{figuretext}
     \end{center}
\end{figure}

\subsection{One-solitons}
Equation (\ref{GNLSgaugesigma1}) admits the one-soliton solutions \cite{L-F}
\begin{equation}\label{onesolitongauge}
u_s(x,t) =  -\frac{2 i e^{i \Sigma+\Theta+2 i \gamma } \sin{\gamma}}{\Delta (e^{2 \Theta+i \gamma }
   +1)},
   \end{equation}
where
\begin{align}
\Sigma  = \frac{\left(t - 4 \Delta^4 x\right) \cos{\gamma}}{2 \Delta^2} + \Sigma_0,
  	\\
\Theta = \frac{\left(t + 4 x \Delta^4\right) \sin{\gamma}}{2 \Delta^2} +  \Theta_0,
   \end{align}
and
$$\gamma \in (0, \pi), \qquad \Delta > 0, \qquad \Sigma_0 \in\R, \qquad \Theta_0 \in \R,$$
are four parameters, see Figure \ref{solitongraphs}. 
In order to relate this family of one-solitons to the above description of traveling waves, we need to find parameters $c_s, k_s, \Omega_s,$ and functions $\theta_s(y), \varphi_s(y)$ such that 
\begin{equation}\label{ustrav}  
  u_s(x,t) = \varphi_s(x-c_st) e^{i(k_s x - \Omega_s t + \theta_s(x - c_s t))}.
\end{equation}
The amplitude of $u_s$ satisfies
$$|u_s|^2 = \frac{2 \sin^2{\gamma}}{\Delta^2 (\cos{\gamma} + \cosh{2 \Theta})}.$$
This yields
\begin{equation}\label{varphis}
\varphi_s(y) = \sqrt{\frac{2 \sin ^2{\gamma}}{\Delta^2 \left(\cos{\gamma} + \cosh \left(4 \Delta^2 y \sin{\gamma} + 2 \Theta_0 \right)\right)}}, \qquad c_s = -\frac{1}{4 \Delta^4}.
\end{equation}
It can be verified that the function $\varphi_s(y)$ given by (\ref{varphis}) satisfies the ODE (\ref{varphiy2V}) provided that $A = B = 0$ and 
\begin{equation}\label{Omegasks}
  k_s = 0, \qquad \Omega_s = -\frac{\cos{\gamma}}{\Delta^2}.
\end{equation}  
Substitution of expression (\ref{varphis}) for $\varphi_s$ into (\ref{thetay}) gives after integration
\begin{align}\label{thetas}
\theta_s(y) =& -\arctan\left(\tan \left(\frac{\gamma}{2}\right) \tanh
   \left(2  \Delta ^2 y \sin{\gamma} + \Theta_0 \right)\right)
   	\\ \nonumber
  & -\cot(\gamma) \left(2 \Delta^2 y \sin{\gamma} + \Theta_0 \right) + \theta_{0s},
\end{align}   
where $\theta_{0s}$ is a constant of integration. We can determine $\theta_{0s}$ by evaluating (\ref{ustrav}) at $x = t = 0$. This yields 

\begin{equation}\label{eitheta0s}
e^{i \theta_{0s}}
= -\frac{i \sqrt{2} e^{i \left(2 \gamma -i \Theta_0+ \Sigma_0 + \arctan\left(\tan \left(\frac{\gamma }{2}\right) \tanh{\Theta_0}\right)+\Theta_0 \cot{\gamma} \right)} \sqrt{\cos{\gamma} +\cosh (2 \Theta_0)}}{1+e^{i \gamma +2
   \Theta_0}}.
\end{equation}

In summary, the one-soliton $u_s$ can be written in the form (\ref{ustrav}) where $c_s, k_s$, $\Omega_s$, $\theta_s(y), \varphi_s(y)$ are given by equations (\ref{varphis})-(\ref{eitheta0s}). This shows how the one-soliton solutions are related to the above traveling-wave analysis. 

\subsection{Peakons}\label{peaksubsec}
\begin{figure}
\begin{center}
    \includegraphics[width=.5\textwidth]{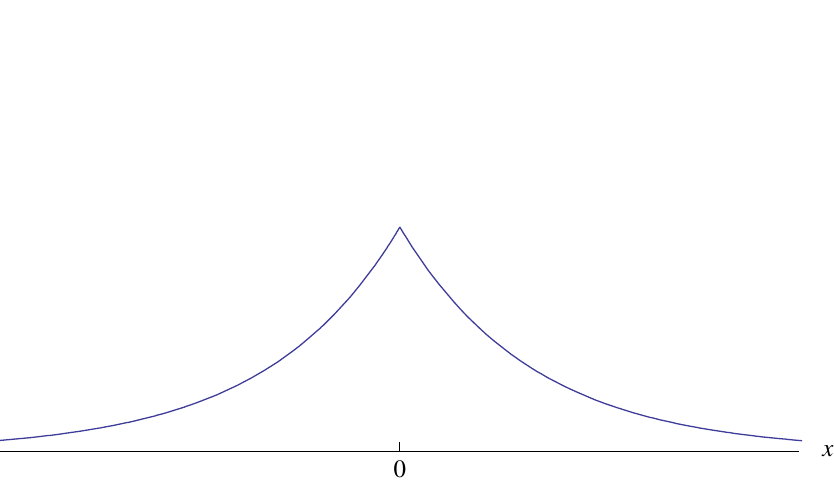} \quad
     \begin{figuretext}\label{peakon.pdf}
       A peakon. The discussion in Subsection \ref{peaksubsec} suggests that equation (\ref{GNLS}) admits no peaked solutions.
     \end{figuretext}
     \end{center}
\end{figure}

The Camassa-Holm equation is well-known for its peakons, which are solitons with a peak at their crest \cite{C-H} (see Figure \ref{peakon.pdf}).
Since equation (\ref{GNLS}) is related to the NLS equation in a similar way that the Camassa-Holm equation is related to KdV, one may wonder if (\ref{GNLS}) also admits some kind of peaked solutions.\footnote{More precisely, we say that a continuous real-valued function $f(x)$ has a peak at $x_0$ if $f$ is
smooth locally on either side of $x_0$ and the left and right derivatives of $f$ at $x_0$ are both finite but not equal. A peaked solution $u(x,t)$ is a solution such that, for each fixed time $t$, $u(\cdot, t)$ has at least one peak. In the case of a complex-valued solution $u(x,t)$, one could allow peaks in either the real or the imaginary part of $u(x,t)$, or in both. }

\begin{figure}
\begin{center}
    \includegraphics[width=.4\textwidth]{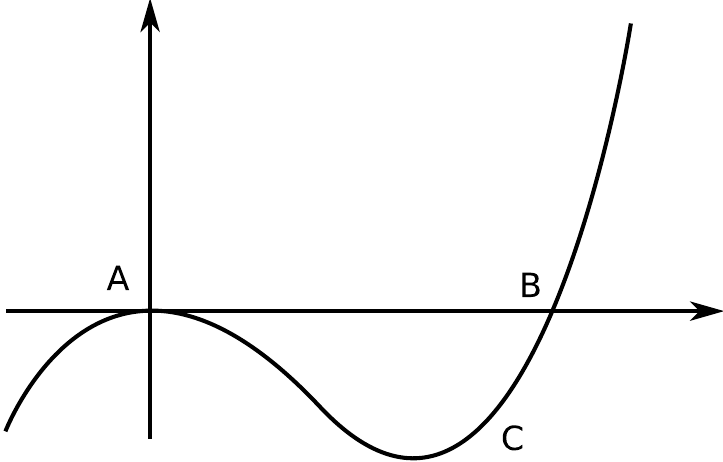} \quad
     \begin{figuretext}\label{potentialwell.pdf}
       An example of a potential well $V(\varphi)$.
          \end{figuretext}
     \end{center}
\end{figure}

Let us first make some general comments on equations of the form (\ref{varphiy2V}) and the occurence of peakons. The ordinary differential equation (\ref{varphiy2V}) can be analyzed qualitatively by regarding $\varphi(y)$ as describing the motion of a ball rolling in a potential well $V(\varphi)$ with $y$ representing time. Using this analogy, peaked solitary waves arise as follows. Consider a potential $V(\varphi)$ with a qualitative shape as in Figure \ref{potentialwell.pdf}. Let the ball at time $y = -\infty$ be located at the crest of the hill at the point $A$. Give it a gentle push so that it starts rolling down the valley. If nothing intervenes, it will climb the opposite side of the valley until it reaches the point $B$ where it turns around; it crosses the valley again and has just enough energy to climb all the way back up to the starting point $A$ as $y \to \infty$. This motion corresponds to a smooth solitary wave solution. However, if we intercept the solution before it reaches point $B$ at some point $C$ at which we let it (after a perfectly elastic collision) rebound, then it will climb up the first hill again and approach its starting position $A$ as $y \to \infty$. The effect of the bounce is to create a peakon singularity at which the derivative $\varphi_y$ jumps, $\varphi_y \to -\varphi_y$. 

Clearly, for most soliton equations there are plenty of ways to intercept solutions in this manner and thus obtain peaked solitons (see \cite{Cheng} for an example of this in the context of fiber optics). However, even though the peaked traveling waves so obtained satisfy the original PDE wherever they are smooth, the singularities in general introduce additional $\delta$-functions so that the PDE is not satisfied in a reasonable weak sense. For most equations this rules out the existence of peaked solitons. On the other hand, if we consider the Camassa-Holm equation 
\begin{equation}\label{CH}
        u_t-u_{txx}+3uu_x =2u_xu_{xx}+uu_{xxx},
\end{equation}
the following happens. For a traveling wave $u(x,t) = \varphi(x - ct)$, integration of (\ref{CH}) yields
\begin{equation}\label{CHtravint}
           -c\varphi+ (c - \varphi)\varphi_{xx} +\frac{3}{2}\varphi^2
           = \frac{1}{2}\varphi_x^2 + \frac{a}{2},
\end{equation}
for some constant of integration $a \in \R$. In going from (\ref{CH}) to (\ref{CHtravint}), the $u_{txx}$ and $uu_{xxx}$ terms have combined to give the term $(c - \varphi)\varphi_{xx}$. Moreover, $(c - \varphi)\varphi_{xx}$ is the only term in (\ref{CHtravint}) which involves a second derivative of $\varphi$. At a point where $\varphi$ has a peak, $\varphi_{xx}$ is proportional to a $\delta$-function. Hence, the singularity-induced $\delta$-function in (\ref{CHtravint}) at a peak appears multiplied by $c - \varphi$, where $c$ is the speed of the wave.  One could therefore hope for a peak to be allowed at a point $x_0$ provided that $c - \varphi(x_0) = 0$, since this factor then kills the $\delta$-function. This is indeed what happens: a peak in a traveling wave of the Camassa-Holm equation is admissible exactly when the height of the peak equals the speed of the wave, i.e. when $\varphi = c$ at the crest cf. \cite{L3}. 

On the contrary, we will now argue that for equation (\ref{GNLS}) the $\delta$-functions produced by the $x$-derivatives of a peakon candidate stand alone and cannot be cancelled by other factors. For simplicity we consider the equivalent, but simpler, equation (\ref{GNLSgaugesigma1}); the argument for (\ref{GNLS}) is analogous. When substituting a traveling-wave solution of the form (\ref{travform}) into equation (\ref{GNLSgaugesigma1}), the terms involving a second derivative of $\varphi$ or $\theta$ can be read off from equation (\ref{travcoupled}). They are
\begin{align*}
-c\varphi_{yy} &\qquad \hbox{in the equation for the real part,}
	\\
 -c\varphi \theta_{yy} & \qquad  \hbox{in the equation for the imaginary part.}
\end{align*}
For the solution to have a peak at $y_0$ at least one of the functions $\varphi_{yy}$ or $\theta_{yy}$ must involve a $\delta$-function at $y_0$. 
The only way for the real and the imaginary parts to vanish simultaneously is therefore that either (i) $c =0$ or (ii) only the function $\theta$ has a peak at $y_0$ and $\varphi(y_0) = \varphi_{yy}(y_0) = 0$. The following observations indicate that neither of these two options is viable. 

Regarding (i) we note that if $c = 0$ then the second equation in (\ref{travcoupled}) becomes
$$-\varphi_y (-\varphi^2 + \Omega) = 0, \qquad y \in \R.$$
By the assumption that the amplitude of a peakon is a continuous function, this shows that the amplitude $\varphi$ is in fact constant. It then follows from the first equation in (\ref{travcoupled}) that $\theta_y$ is also a constant function given by
$$\theta_y = \frac{k \varphi^2-k \Omega + 1}{\Omega - \varphi^2}.$$
Thus, no peakons arise when $c = 0$.

Regarding (ii) we note that a vanishing amplitudeÊ$\varphi(y_0) = 0$ at $y_0$ implies that the phase $\theta(y_0)$ is undetermined a priori. Of course $\theta$ may contain peaks (or jumps for that matter) in any intervals where $\varphi = 0$, but these peaks are unobservable and are just a consequence of the breakdown of the polar coordinate system at the origin. As soon as $\varphi$ becomes nonzero the value of $\theta_y$ is fixed by the equations in (\ref{travcoupled}) and cannot jump. Hence, neither of the options (i) or (ii) leads to existence of peakons. 

The above discussion indicates that (\ref{GNLSgaugesigma1}) admits no peaked solutions within the class of traveling waves of the form (\ref{travform}). Although one could imagine that there exist peaked solutions of a more general type, similar arguments as above applied directly to equation (\ref{GNLSgaugesigma1}) make this unlikely (since $u_{tx}$ is the only term in the equation involving two derivatives, there is no other term to balance the $\delta$-function generated by $u_{tx}$ at the peak). In conclusion, it appears that equation (\ref{GNLS}) admits no peakons which are weak solutions in a reasonable sense.

\bibliography{is}

\end{document}